\documentclass[runningheads]{llncs}
\usepackage{graphicx}
\usepackage{tikz}
\usepackage{comment}
\usepackage{booktabs}
\usepackage{cite}
\usepackage{paralist}
\usepackage{fancyvrb}
\usepackage{subfigure}
\usepackage{siunitx}
\usepackage{dcolumn}
\usepackage{xcolor}
\usepackage{colortbl}
\usepackage{xspace}

\newcommand{\nop}[1]{{}} 
\usepackage{xcolor}
\usepackage{marginnote} 

\renewcommand{\marginnote}[2][]{}

\usepackage[colorlinks=true,urlcolor=blue,linkcolor = blue,  anchorcolor = blue]{hyperref}
\newcommand{\webref}[2]{\href{#1}{#2}} 
\usepackage{marginnote} 

\newcommand{\vspacesml}{\vspace*{-2mm}}

\newcommand{\mysqldiff}{\texttt{mysqldiff}\xspace}

\begin{document}
\title{Replicability and Reproducibility of a Schema Evolution Study in Embedded Databases}
%
\titlerunning{Examining Replicability and Reproducibility}
%
\author{Dimitri Braininger\inst{1}
\and
Wolfgang Mauerer\inst{1,2}
\and
Stefanie Scherzinger\inst{3}
}
\authorrunning{D.\ Braininger, W.\ Mauerer, S.\ Scherzinger}
%
\institute{Technical University of Applied Sciences Regensburg, Regensburg, Germany
\email{dimitri.braininger@st.oth-regensburg.de}
\and
Siemens AG, Corporate Research, Munich 
\email{wolfgang.mauerer@othr.de}\\
\and
University of Passau, Germany
\email{stefanie.scherzinger@uni-passau.de}}
\maketitle              
\begin{abstract}

Ascertaining the feasibility of independent falsification or repetition of published results is vital to the scientific process, and replication or reproduction experiments are routinely
performed in many disciplines. Unfortunately, such studies are only scarcely available in database
research, with few papers dedicated to re-evaluating published results. 
In this paper, we conduct a case study on replicating and reproducing a study on schema evolution
in embedded databases. 
We can exactly repeat the outcome for one out of four database applications studied, and come close in two further cases.
By reporting results, efforts, and obstacles encountered, 
we hope to
increase appreciation for the substantial efforts required to ensure reproducibility.
By discussing minutiae details required to ascertain  reproducible work,
we argue that such important, but often ignored aspects of scientific work should receive more credit in the evaluation of future
research.

\end{abstract}

\section{Introduction}

Experiments are at the heart of the scientific process. According to the ACM reproducibility guidelines (see ``\webref{https://www.acm.org/publications/policies/artifact-review-badging}{ACM review and badging}'', hyperlink available in the PDF), experiments are expected to be \emph{repeatable}:  Essentially, the same team with the same experimental setup can reliably 
achieve identical results in subsequent trials.
Moreover, experiments should be \emph{replicable}, so that using the same experimental setup operated by a different team achieves the same results.
Ideally, experiments are even \emph{reproducible}, and a different team with a different experimental setup can confirm the results.

Such properties are acknowledged to be fundamental, but
reproducibility is far from universally permeating most published research.   This discrepancy has become an academic topic of debate, and dedicated research evaluates the (oftentimes wanting) state of affairs in computer science research in general
(see, e.g., Refs.~\cite{10.1145/3385658.3385668,pawlik2019,DBLP:journals/sigmod/ManolescuAADMPSSZS08,10.1145/2812803}), but also in data management research\footnote{Such as in
the VLDB (``\webref{https://vldb-repro.com/}{pVLDB Reproducibility}'') and SIGMOD communities (``\webref{http://db-reproducibility.seas.harvard.edu/}{ACM SIGMOD 2019 Reproducibility}'',
clickable links available in PDF).}.

In this paper, we examine the state of replicability, and efforts required to achieve reproducibility, for an empirical case study on schema evolution in embedded databases by  S.\ Wu and I.\ Neamtiu~\cite{original_work} that predates the aforementioned discussions. %
There is a long-standing tradition of schema evolution case studies in real-world database applications, e.g.,~\cite{Curino08schemaevolution,Qiu:2013:EAC:2491411.2491431,DBLP:journals/infsof/Sjoberg93,DBLP:conf/er/VassiliadisZS15}.
It used to be difficult to get access to  real-world database applications for study, so earlier studies are generally conducted on closed-source systems, for instance~\cite{DBLP:journals/infsof/Sjoberg93}.
Yet the proliferation of open source software, and the access to code repositories (e.g., \webref{https://github.com/}{GitHub}) enables a whole new line of research on open source application code~\cite{Bird:2015:ASA:2886235}.
Most schema evolution studies focus on applications backed by relational database management systems, typically tracking the growth of the schema (counting the number of tables and their columns), and the distribution of \emph{schema modification operations} (a term coined by Curino et al.\ in~\cite{curino_prism2008}).

The authors in the original case study are the first to focus on an important subfamily of database products, namely that of \emph{embedded} (and therefore serverless) databases, such as \webref{https://www.sqlite.org/index.html}{SQLite}. 
%
While there are independent schema evolution studies targeting the same open source projects, such as  \webref{https://www.mediawiki.org/wiki/MediaWiki}{MediaWiki} (the software powering Wikipedia),
they consider different time frames (such as 4.5 years in~\cite{Curino08schemaevolution} and 10 years in~\cite{Qiu:2013:EAC:2491411.2491431}), and implement different methodologies. This even leads to  partly  contradictory results. However, a dedicated replicability and reproducibility study has not yet been conducted so far.

\smallskip
\noindent
\emph{Contributions.}
This paper makes the following contributions:
\begin{compactitem}
\item We conduct a replicability and reproducibility study on a well-received, published paper on schema evolution~\cite{original_work}. While there is a long history of schema evolution case studies, to the best of our knowledge, ours is the first effort to ascertain published results on this class of publications.

\item Our study is mainly based on the information provided in the original paper. However, we were also provided (incomplete) code artefacts by the authors of the original study.
This blurs the line between conducting a replicability and reproducibility study. For simplification, we restrict ourselves to the term \emph{reproducibility} in the remainder of this paper.

\item We carefully re-engineer the authors' experiments and present our results. Overall, we achieve a high degree of accordance, albeit at the expense of substantial manual effort.
For one out of four applications studied in~\cite{original_work}, we even obtain identical results. 
We document and discuss where our numbers agree, and where they deviate.

\item We lay out which instructions  were helpful, and which left too much leeway.

\item We discuss the threats to the validity of our results (e.g., where we may have erred), and contrast this with the original threats stated in~\cite{original_work}. Doing so, we re-calibrate the level of risk involved with each originally reported threat. 
\end{compactitem}
Our experience underlines that achieving full reproducibility remains a challenge even with well-designed, well-documented studies, and requires considerable extra effort. We feel that such efforts are not yet universally appreciated, albeit it is in our joint interest that research become reproducible.

\paragraph{Structure.} 
The remainder of this paper is organized as follows.
We next summarize the original study. Section~\ref{sec:methodology} states our methodology. Section~\ref{results} describes the main part of the reproduction work, as well as the detailed results.
Section~\ref{discussion} discusses the overall results, followed by Section~\ref{threats} with a description of  threats to validity. Finally, Section~\ref{related} focuses on related work. Section~\ref{conclusion} concludes. 

\section{Original Study}
\label{orig_study}

We briefly summarize the original study. Neamtiu et al.\ analyze four database applications, all of which are based
on SQLite, and provide public development histories
by virtue of being available as open source software (OSS):
\webref{https://textbrowser.github.io/biblioteq/}{\emph{BiblioteQ}}, \webref{https://www.monotone.ca/}{\emph{Monotone}}, \webref{https://www.mozilla.org/de/firefox/}{\emph{Mozilla Firefox}}, and \webref{https://www.vienna-rss.com/}{\emph{Vienna}}:
\begin{compactitem}

\item \emph{BiblioteQ} (C\texttt{++}), analyzed in the time frame 03/15/2008--02/19/2010, is a library management system. 

\item \emph{Monotone} (C\texttt{++}), analyzed in the time frame  04/06/2003--06/13/2010, is a distributed version control system.

\item \emph{Mozilla Firefox} (C, C\texttt{++}), analyzed in the time frame  10/02/2004--11/21/2008, is a popular web browser. 

\item \emph{Vienna} (Objective-C), analyzed in the time frame 06/29/2005--09/03/2010, is an RSS newsreader for MacOS. 
\end{compactitem}

The original study uses a custom data processing pipeline for retrieving the source code history,
extracting schema declarations embedded in  application  code, and computing
differences between schema revisions. 
Extracting schema declarations
requires careful engineering:
Figure~\ref{subfig:cpp} shows a  CREATE TABLE statement embedded in the program code as a multi-line string constant.

\begin{figure}[t]
    \flushleft
    
\begin{SaveVerbatim}[]{VerbEnv1}
res = logged_sqlite3_exec(sql, "CREATE TABLE file_deltas\n"
 "\t(\n"
 "\tid not null,    -- strong hash of file contents\n"
 "\tbase not null,  -- joins with files.id or file_deltas.id\n"
 "\tdelta not null, -- compressed [...]\n"
 "\tunique(id, base)\n"
 "\t)", NULL, NULL, errmsg);
\end{SaveVerbatim}    

\begin{SaveVerbatim}[]{VerbEnv2}
CREATE TABLE file_deltas
(
 id integer not null, 
 base integer not null, 
 delta integer not null, 
 unique(id, base)
);  
\end{SaveVerbatim}  

\subfigure[Excerpt from the C\texttt{++} code in \emph{Monotone}.]{ \label{subfig:cpp}
\fbox{\BUseVerbatim[fontsize=\scriptsize]{VerbEnv1}}
}%
\subfigure[Extracted stmt.]{\label{subfig:parsed}
\fbox{\BUseVerbatim[fontsize=\scriptsize]{VerbEnv2}}
}

    \caption{(a) A CREATE TABLE statement,
    embedded as string constants within  \emph{Monotone} C\texttt{++} code (source can be inspected \webref{https://github.com/brdd3v/repro/blob/master/monotone/input/cc/schema_migration_026.cc}{online}, ``{\tt [...]}'' denotes a shortened comment). The statement must be automatically parsed and translated to the MySQL dialect (b).}
    \label{fig:create_table_embedded}
\end{figure}

We compare different schema versions with \webref{https://github.com/aspiers/mysqldiff}{\mysqldiff}
(version 0.30), a utility to derive schema modification operations (SMOs)  that transform a predecessor schema into the successor schema.
 \mysqldiff only handles MySQL schema declarations, but SQLite uses a custom SQL dialect\footnote{The SQL dialects reference at \url{https://en.wikibooks.org/wiki/SQL_Dialects_Reference} illustrates the richness of proprietary language constructs.}. 
%
For instance, let us again consider the code example from Figure~\ref{subfig:cpp}. The extracted CREATE TABLE statement is  shown in Figure~\ref{subfig:parsed}. Note that the original statement does not declare attribute types, which is permissible when using SQLite. Since MySQL requires all attributes to be typed, we add a default attribute type in preparation for processing the schemas with \mysqldiff.

\mysqldiff generates SMOs for creating or dropping a table, adding or removing a table column, and  changing the type or initial value of a column. It also recognizes changes to the table primary key.
With this sequence of SMOs, the predecessor schema can be transformed into its successor schema.
Further SMOs, such as renaming a table or an attribute, cannot be reliably derived based on automated analysis alone, and would require sophisticated schema matching and mapping solutions~\cite{Bellahsene:2011:SMM:1972526}.

The statistics in the study by Neamtiu et al.\ derive from
 \mysqldiff results; Table~\ref{fig_table} provides the number of SMOs for each project.
%
Studies on schema evolution in server-based (non-embedded) DBMS, especially~\cite{Qiu:2013:EAC:2491411.2491431}, show that attribute type changes are frequent in many projects. In the study by Neamtiu et al., this holds only for \emph{BiblioteQ},
so no type changes were recorded for the other projects. This is a finding that we will revisit at a later point.
The original study finds that the shares of CREATE TABLE and ADD COLUMN SMOs are comparable to the observations of related studies on schema evolution in non-embedded DBMS. The observation that changes to initial values and primary keys are uncommon has also been observed in the later study of Qiu et al.~\cite{Qiu:2013:EAC:2491411.2491431}.

\nop{
\begin{table}[t]
\caption{Evolution time frames and schema change details (as absolute numbers and percentages) given in the original study~\cite{original_work}.} \label{fig_table}
\includegraphics[width=\textwidth]{img/orig_work_table.png}
\end{table}
}
{

\begingroup
\setlength{\tabcolsep}{1.87pt}
\begin{table}[t]
\centering
\caption{Evolution time frames and schema change details (as absolute numbers and percentages) given in the original study~\cite{original_work}.} 
\label{fig_table}

{\scriptsize
\newcommand{\spacer}{\hspace*{0.3em}}
\newcommand{\capcorr}{-0.15em}
\newcommand{\cc}{\columncolor{gray}}
\newcolumntype{d}[1]{D{.}{.}{#1}}
\begin{tabular}{l rr>{(}d{2.1}@{\%) } rr>{(}d{2.1}@{\%) } 
c
                  rr>{(}d{2.1}@{\%) } rr>{(}d{2.1}@{\%) }      
                  rr>{(}d{2.1}@{\%) } rr>{(}d{2.1}@{\%) } rr>{(}d{2.1}@{\%) } }
\toprule
\vbox to0pt{\hbox to0pt{\textbf{App}}} &  \multicolumn{6}{c}{\textbf{Table changes}} & \spacer &  \multicolumn{15}{c}{\textbf{Attribute changes}}\\
\cline{2-7}\cline{9-23}\\[-0.7em]
 & \multicolumn{3}{c}{{\shortstack{\strut CREATE\\[\capcorr]TABLE}}} & 
   \multicolumn{3}{c}{{\shortstack{\strut DROP\\[\capcorr]TABLE}}}   & \spacer &
   \multicolumn{3}{c}{{{\shortstack{\strut ADD\\[\capcorr]COLUMN}}}} &
   \multicolumn{3}{c}{{\shortstack{\strut DROP\\[\capcorr]COLUMN}}}  &
   \multicolumn{3}{c}{{\shortstack{\strut Type\\[\capcorr]change}}}  &
   \multicolumn{3}{c}{{\shortstack{\strut Init\\[\capcorr]change}}}  &
   \multicolumn{3}{c}{{\shortstack{\strut Key\\[\capcorr]change}}}\\ 
\midrule
\textbf{Firefox}    && 5  & 4.2  && 26 & 21.7 & \spacer && 57  & 47.5 && 28 & 23.3 && 0  & 0.0  && 3 & 2.5 && 1 & 0.7 \\
\textbf{Monotone}   && 11 & 20.4 && 17 & 31.5 & \spacer && 14  & 25.9 && 10 & 18.5 && 0  & 0.0  && 0 & 0.0 && 2 & 3.7 \\
\textbf{BiblioteQ}  && 4  & 2.6  && 8  & 5.2  & \spacer && 27  & 17.5 && 28 & 18.2 && 83 & 53.9 && 0 & 0.0 && 4 & 2.6 \\
\textbf{Vienna}     && 1  & 7.1  && 0  & 0.0  & \spacer && 13  & 92.9 && 0  & 0.0  && 0  & 0.0  && 0 & 0.0 && 0 & 0.0 \\
\midrule
\textbf{Total}      && 21 & 6.1  && 51 & 14.9 & \spacer && 111 & 32.5 && 66 & 19.3 && 83 & 24.3 && 3 & 0.9 && 7 & 2.0 \\
\bottomrule
\end{tabular}
}

\end{table}
\endgroup
}

\section{Methodology of this Study} 
\label{sec:methodology}
We conducted our reproducibility study as follows. 
Our code, as well as material made available to us by the original authors, is available on \href{http://doi.org/10.5281/zenodo.4012776}{Zenodo} (doi.org/10.5281/zenodo.4012776) to ascertain long-term availability. In particular, we publish all interim results computed by our analysis scripts (such as the extracted schemas and the results of schema comparison), for transparency. 

We started with identifying the source code repositories for the four database applications, based on the information given in the original paper.
Like in the original work, we wrote a script to extract the database schema declarations embedded in the source code.
For  \emph{Vienna}, the authors provided us with a partial script that could not be directly made to work (caused by minor syntactic issues, and some missing components), and was therefore re-implemented by us in Python.
For all other projects, we had no such templates.


The original study used \mysqldiff version~0.30 to compare successive schema declarations. However, we used the newer version~0.60, 
since the output is more succinct and also more convenient to parse. A further reason for abandoning the legacy version is that it sometimes recognizes redundant schema modification operations  (as we also discuss in Section~\ref{threats}).

Further, the  pairwise comparison of schema versions using \mysqldiff is not very robust: A table declaration that is missing in one version (e.g., due to a parsing problem), and then re-appears later, is recognized as first dropping and later re-creating this table. 
This problem was pointed out in the original study,
and will also be revisited in Section~\ref{threats}.

As a summarizing metric,
we compute the difference in percentage across all SMOs observed
as
\begin{displaymath}
\frac{\sum_{SMO\; s} \left | p(s) - r(s)\right |}{P},
\end{displaymath}
where 
$p(s)$ is the number of changes for SMO~$s$ reported in the original publication and
$r(s)$ is the number of changes for SMO~$s$ identified in our reproducibility study. Further,~$P$ is the total number of changes in the project reported in~\cite{original_work}.

\section{Results} \label{results}

\paragraph{Vienna.}

For \emph{Vienna}, the authors made their raw input data available to us, so we could apply our script on the exact same data, with the exact same results.

We further attempted to locate the raw input data ourselves, based only on information provided in the original study.
Unfortunately, the original Sourceforge repository no longer exists,  the project is now hosted on \webref{https://github.com/ViennaRSS/vienna-rss}{GitHub}. From there, we obtained fewer files than expected. 
Thus, searching for the raw input data based on the information in the paper alone would have led to a different baseline, yet the analysis still yields the same results as listed in Table~\ref{fig_table}.

\paragraph{Monotone.}

For \emph{Monotone}, the original paper states that the study was conducted on 
48 archives available from the project website.
However, we have reason to believe that only 41 versions were chosen (specifically, versions 0.1, 0.2, and also from 0.10 up to and including 0.48),
 based 
 on the list of available archives, as well as comments within the material that we obtained from the authors.

\nop{
From e-mail from 19-APR-2020:
[Monotone]
Ich habe früher unter den empfangenen Daten eine Textdatei gefunden, in der 41 Versionen aufgelistet waren:
monotone-0.1:        11
monotone-0.2:        11
monotone-0.10:        14
...
monotone-0.48:        16
Aufgrund der Erwähnung im Paper von 48 Versionen ging ich trotzdem davon aus, dass es 48 Versionen analysiert waren.
Nach der Überprüfung habe ich in einer Datei  Erwähnung von 41 Versionen gefunden.
}

Moreover, it is not exactly clear from which files to extract schema declarations: In the initial versions of \emph{Monotone}, database schemas are only declared in files with suffix \verb!.sql!. Later, database schemas are also embedded within C\texttt{++} files (starting with version~10). 
We therefore explored two approaches, where we
(1) consider \emph{only} schemas declared in \verb!.sql!-suffixed files, and
(2) also consider schemas embedded within the program code.

\begin{figure}[tb]
\centering
\scalebox{1}{\input{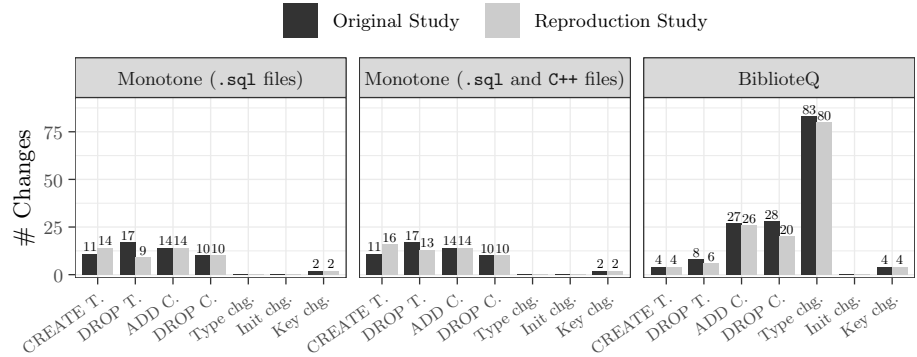}}
\vspace*{-2em}
\caption{Comparing the number of schema changes for \emph{Monotone} and \emph{BiblioteQ}.}
\label{fig_combo}
\end{figure}

Figure~\ref{fig_combo} visualizes the results for both approaches. For each type of SMO analyzed, we compare the number of changes reported in the original study with the number of changes determined by us.
Overall, our results come close.
As pointed out in Section~\ref{sec:methodology},
problems in parsing SQL statements embedded in program code
lead to falsely recognizing tables as dropped and later re-introduced.
We suspect that this effect causes the discrepancies observed for CREATE and DROP  TABLE statements.


\nop{
It was not a script, but a text file "Monotone/smo_count" among the data.
Its contents:
monotone-0.45 : 6
monotone-0.18 : 1
monotone-0.32 : 2
monotone-0.46 : 2
monotone-0.17 : 3
monotone-0.18 : 1
monotone-0.15 : 9
monotone-0.26 : 6
monotone-0.10 : 19
monotone-0.24 : 1
monotone-0.30 : 4
The line "monotone-0.18 : 1" is contained twice.
If we sum the values, we get 54, which is the total number of SMOs in the Monotone project in the original work.
I do not know whether it is possible to rely on this data.
I did not add this file to the repository, because didn't know if this file was important.
}


\nop{
\begin{figure}[tb]
\centering
\scalebox{0.90}{\input{img/monotone.tex}}
\vspacesml
\caption{Comparing the number of schema changes for \emph{Monotone}.}
\label{fig_monotone}
\end{figure}
}

\paragraph{BiblioteQ.}

At the time when the original study was performed on \emph{BiblioteQ}, all schema declarations were contained in files with suffix \verb!.sql! (this has meanwhile changed).
Schema declarations do thus not have to be laboriously parsed from strings embedded in the application source code.
MySQL, SQLite, and PostgreSQL were supported as alternative backends.
In particular, SQLite was initially not supported, but was introduced with revision~35, while the original study spans the time frame from the very beginning of the project (see Section~\ref{orig_study}). 
Unfortunately, the original study does not discuss this issue.

We suspect that up to revision~35, the schema declarations of MySQL were analyzed, and only from then on for SQLite.%
\footnote{Revision~16 only changes the MySQL schema declaration, and the original study reports a schema change in this revision.  A peak in schema changes is reported for revision~35 (see Table~\ref{tab_biblioteq_refm}), as switching from MySQL to SQLite schema declarations causes \mysqldiff to recognize type changes.
Since revision~35 only adds support for SQLite, with no schema changes for MySQL or PostgreSQL, we conclude that starting with revision~35, the authors analyzed the SQLite schema.}
%
The high number of type changes reported for \emph{BiblioteQ} 
may thus be 
overemphasized---the switch causes half the reported type changes. However, this still leaves a significant number of type changes for BiblioteQ, compared to the other 
projects (see Table~\ref{fig_table}). \marginnote{We need to phrase this very carefully!}


\nop{
\begin{figure}[tb]
\centering
\scalebox{0.90}{
\begin{tikzpicture}[x=1pt,y=1pt]
\definecolor{fillColor}{RGB}{255,255,255}
\path[use as bounding box,fill=fillColor,fill opacity=0.00] (0,0) rectangle (347.12,138.85);
\begin{scope}
\path[clip] (  0.00,  0.00) rectangle (347.12,138.85);
\definecolor{drawColor}{RGB}{255,255,255}
\definecolor{fillColor}{RGB}{255,255,255}

\path[draw=drawColor,line width= 0.5pt,line join=round,line cap=round,fill=fillColor] ( -0.00,  0.00) rectangle (347.12,138.85);
\end{scope}
\begin{scope}
\path[clip] ( 23.83, 31.81) rectangle (240.29,138.85);
\definecolor{fillColor}{RGB}{255,255,255}

\path[fill=fillColor] ( 23.83, 31.81) rectangle (240.29,138.85);
\definecolor{drawColor}{gray}{0.92}

\path[draw=drawColor,line width= 0.3pt,line join=round] ( 23.83, 50.16) --
	(240.29, 50.16);

\path[draw=drawColor,line width= 0.3pt,line join=round] ( 23.83, 77.72) --
	(240.29, 77.72);

\path[draw=drawColor,line width= 0.3pt,line join=round] ( 23.83,105.27) --
	(240.29,105.27);

\path[draw=drawColor,line width= 0.3pt,line join=round] ( 23.83,132.83) --
	(240.29,132.83);

\path[draw=drawColor,line width= 0.5pt,line join=round] ( 23.83, 36.38) --
	(240.29, 36.38);

\path[draw=drawColor,line width= 0.5pt,line join=round] ( 23.83, 63.94) --
	(240.29, 63.94);

\path[draw=drawColor,line width= 0.5pt,line join=round] ( 23.83, 91.50) --
	(240.29, 91.50);

\path[draw=drawColor,line width= 0.5pt,line join=round] ( 23.83,119.05) --
	(240.29,119.05);

\path[draw=drawColor,line width= 0.5pt,line join=round] ( 41.87, 31.81) --
	( 41.87,138.85);

\path[draw=drawColor,line width= 0.5pt,line join=round] ( 71.93, 31.81) --
	( 71.93,138.85);

\path[draw=drawColor,line width= 0.5pt,line join=round] (101.99, 31.81) --
	(101.99,138.85);

\path[draw=drawColor,line width= 0.5pt,line join=round] (132.06, 31.81) --
	(132.06,138.85);

\path[draw=drawColor,line width= 0.5pt,line join=round] (162.12, 31.81) --
	(162.12,138.85);

\path[draw=drawColor,line width= 0.5pt,line join=round] (192.18, 31.81) --
	(192.18,138.85);

\path[draw=drawColor,line width= 0.5pt,line join=round] (222.25, 31.81) --
	(222.25,138.85);
\definecolor{fillColor}{gray}{0.20}

\path[fill=fillColor] ( 28.34, 36.38) rectangle ( 41.87, 40.79);

\path[fill=fillColor] ( 58.40, 36.38) rectangle ( 71.93, 45.20);

\path[fill=fillColor] ( 88.47, 36.38) rectangle (101.99, 66.14);

\path[fill=fillColor] (118.53, 36.38) rectangle (132.06, 67.25);

\path[fill=fillColor] (148.59, 36.38) rectangle (162.12,127.87);

\path[fill=fillColor] (178.66, 36.38) rectangle (192.18, 36.38);

\path[fill=fillColor] (208.72, 36.38) rectangle (222.25, 40.79);
\definecolor{fillColor}{gray}{0.80}

\path[fill=fillColor] ( 41.87, 36.38) rectangle ( 55.40, 40.79);

\path[fill=fillColor] ( 71.93, 36.38) rectangle ( 85.46, 43.00);

\path[fill=fillColor] (101.99, 36.38) rectangle (115.52, 65.04);

\path[fill=fillColor] (132.06, 36.38) rectangle (145.59, 58.43);

\path[fill=fillColor] (162.12, 36.38) rectangle (175.65,124.56);

\path[fill=fillColor] (192.18, 36.38) rectangle (205.71, 36.38);

\path[fill=fillColor] (222.25, 36.38) rectangle (235.78, 40.79);
\definecolor{drawColor}{RGB}{0,0,0}

\node[text=drawColor,anchor=base,inner sep=0pt, outer sep=0pt, scale=  0.63] at ( 35.85, 41.87) {4};

\node[text=drawColor,anchor=base,inner sep=0pt, outer sep=0pt, scale=  0.63] at ( 65.92, 46.28) {8};

\node[text=drawColor,anchor=base,inner sep=0pt, outer sep=0pt, scale=  0.63] at ( 95.98, 67.22) {27};

\node[text=drawColor,anchor=base,inner sep=0pt, outer sep=0pt, scale=  0.63] at (126.04, 68.32) {28};

\node[text=drawColor,anchor=base,inner sep=0pt, outer sep=0pt, scale=  0.63] at (156.11,128.95) {83};

\node[text=drawColor,anchor=base,inner sep=0pt, outer sep=0pt, scale=  0.63] at (216.24, 41.87) {4};

\node[text=drawColor,anchor=base,inner sep=0pt, outer sep=0pt, scale=  0.63] at ( 47.88, 41.87) {4};

\node[text=drawColor,anchor=base,inner sep=0pt, outer sep=0pt, scale=  0.63] at ( 77.94, 44.07) {6};

\node[text=drawColor,anchor=base,inner sep=0pt, outer sep=0pt, scale=  0.63] at (108.01, 66.12) {26};

\node[text=drawColor,anchor=base,inner sep=0pt, outer sep=0pt, scale=  0.63] at (138.07, 59.51) {20};

\node[text=drawColor,anchor=base,inner sep=0pt, outer sep=0pt, scale=  0.63] at (168.13,125.64) {80};

\node[text=drawColor,anchor=base,inner sep=0pt, outer sep=0pt, scale=  0.63] at (228.26, 41.87) {4};
\definecolor{drawColor}{gray}{0.20}

\path[draw=drawColor,line width= 0.5pt,line join=round,line cap=round] ( 23.83, 31.81) rectangle (240.29,138.85);
\end{scope}
\begin{scope}
\path[clip] (  0.00,  0.00) rectangle (347.12,138.85);
\definecolor{drawColor}{gray}{0.30}

\node[text=drawColor,anchor=base east,inner sep=0pt, outer sep=0pt, scale=  0.80] at ( 19.33, 33.63) {0};

\node[text=drawColor,anchor=base east,inner sep=0pt, outer sep=0pt, scale=  0.80] at ( 19.33, 61.18) {25};

\node[text=drawColor,anchor=base east,inner sep=0pt, outer sep=0pt, scale=  0.80] at ( 19.33, 88.74) {50};

\node[text=drawColor,anchor=base east,inner sep=0pt, outer sep=0pt, scale=  0.80] at ( 19.33,116.30) {75};
\end{scope}
\begin{scope}
\path[clip] (  0.00,  0.00) rectangle (347.12,138.85);
\definecolor{drawColor}{gray}{0.20}

\path[draw=drawColor,line width= 0.5pt,line join=round] ( 21.33, 36.38) --
	( 23.83, 36.38);

\path[draw=drawColor,line width= 0.5pt,line join=round] ( 21.33, 63.94) --
	( 23.83, 63.94);

\path[draw=drawColor,line width= 0.5pt,line join=round] ( 21.33, 91.50) --
	( 23.83, 91.50);

\path[draw=drawColor,line width= 0.5pt,line join=round] ( 21.33,119.05) --
	( 23.83,119.05);
\end{scope}
\begin{scope}
\path[clip] (  0.00,  0.00) rectangle (347.12,138.85);
\definecolor{drawColor}{gray}{0.20}

\path[draw=drawColor,line width= 0.5pt,line join=round] ( 41.87, 29.31) --
	( 41.87, 31.81);

\path[draw=drawColor,line width= 0.5pt,line join=round] ( 71.93, 29.31) --
	( 71.93, 31.81);

\path[draw=drawColor,line width= 0.5pt,line join=round] (101.99, 29.31) --
	(101.99, 31.81);

\path[draw=drawColor,line width= 0.5pt,line join=round] (132.06, 29.31) --
	(132.06, 31.81);

\path[draw=drawColor,line width= 0.5pt,line join=round] (162.12, 29.31) --
	(162.12, 31.81);

\path[draw=drawColor,line width= 0.5pt,line join=round] (192.18, 29.31) --
	(192.18, 31.81);

\path[draw=drawColor,line width= 0.5pt,line join=round] (222.25, 29.31) --
	(222.25, 31.81);
\end{scope}
\begin{scope}
\path[clip] (  0.00,  0.00) rectangle (347.12,138.85);
\definecolor{drawColor}{gray}{0.30}

\node[text=drawColor,rotate= 35.00,anchor=base east,inner sep=0pt, outer sep=0pt, scale=  0.70] at ( 44.63, 23.36) {CREATE T.};

\node[text=drawColor,rotate= 35.00,anchor=base east,inner sep=0pt, outer sep=0pt, scale=  0.70] at ( 74.70, 23.36) {DROP T.};

\node[text=drawColor,rotate= 35.00,anchor=base east,inner sep=0pt, outer sep=0pt, scale=  0.70] at (104.76, 23.36) {ADD C.};

\node[text=drawColor,rotate= 35.00,anchor=base east,inner sep=0pt, outer sep=0pt, scale=  0.70] at (134.82, 23.36) {DROP C.};

\node[text=drawColor,rotate= 35.00,anchor=base east,inner sep=0pt, outer sep=0pt, scale=  0.70] at (164.89, 23.36) {Type chg.};

\node[text=drawColor,rotate= 35.00,anchor=base east,inner sep=0pt, outer sep=0pt, scale=  0.70] at (194.95, 23.36) {Init chg.};

\node[text=drawColor,rotate= 35.00,anchor=base east,inner sep=0pt, outer sep=0pt, scale=  0.70] at (225.01, 23.36) {Key chg.};
\end{scope}
\begin{scope}
\path[clip] (  0.00,  0.00) rectangle (347.12,138.85);
\definecolor{drawColor}{RGB}{0,0,0}

\node[text=drawColor,rotate= 90.00,anchor=base,inner sep=0pt, outer sep=0pt, scale=  1.00] at (  6.89, 85.33) {\# Changes};
\end{scope}
\begin{scope}
\path[clip] (  0.00,  0.00) rectangle (347.12,138.85);
\definecolor{fillColor}{RGB}{255,255,255}

\path[fill=fillColor] (250.29, 63.37) rectangle (347.12,107.28);
\end{scope}
\begin{scope}
\path[clip] (  0.00,  0.00) rectangle (347.12,138.85);
\definecolor{fillColor}{RGB}{255,255,255}

\path[fill=fillColor] (255.29, 82.83) rectangle (269.74, 97.28);
\end{scope}
\begin{scope}
\path[clip] (  0.00,  0.00) rectangle (347.12,138.85);
\definecolor{fillColor}{gray}{0.20}

\path[fill=fillColor] (256.00, 83.54) rectangle (269.03, 96.57);
\end{scope}
\begin{scope}
\path[clip] (  0.00,  0.00) rectangle (347.12,138.85);
\definecolor{fillColor}{RGB}{255,255,255}

\path[fill=fillColor] (255.29, 68.37) rectangle (269.74, 82.83);
\end{scope}
\begin{scope}
\path[clip] (  0.00,  0.00) rectangle (347.12,138.85);
\definecolor{fillColor}{gray}{0.80}

\path[fill=fillColor] (256.00, 69.09) rectangle (269.03, 82.12);
\end{scope}
\begin{scope}
\path[clip] (  0.00,  0.00) rectangle (347.12,138.85);
\definecolor{drawColor}{RGB}{0,0,0}

\node[text=drawColor,anchor=base west,inner sep=0pt, outer sep=0pt, scale=  0.80] at (274.74, 87.30) {original study};
\end{scope}
\begin{scope}
\path[clip] (  0.00,  0.00) rectangle (347.12,138.85);
\definecolor{drawColor}{RGB}{0,0,0}

\node[text=drawColor,anchor=base west,inner sep=0pt, outer sep=0pt, scale=  0.80] at (274.74, 72.85) {repro. study};
\end{scope}
\end{tikzpicture}}
\vspacesml
\caption{Comparing the number of schema changes for \emph{BiblioteQ}.}
\label{fig_biblioteq}
\end{figure}
}

\begin{table}[t]
\centering
\caption{Pairwise comparison of schema versions, and the number of changes w.r.t.\ the previous version. Stating the number of changes reported in the original paper (\#C, original), the number of changes identified in our reproducibility study (\#C, repro), as well as the absolute difference (diff), for \textit{BiblioteQ}.}
\label{tab_biblioteq_refm}

{
\footnotesize
\resizebox{\textwidth}{!}{%
\begin{tabular}{lrrrrrrrrrrrrrrrrrrr}
\toprule
\bf Revision & 4 & 5 & 11 & 16 & 24 & 35 & 44 & 52 & 80 & 81 & 101 & 102 & 115 & 116 & 154 & 233 & 236 & 285 & \bf Total \\
\midrule
\bf \#C,original & 1 & 1 & 1 & 1  & 20 & 50 & 5 & 25 & 1 & 8 & 12 & 12 & 1 & 5 & 1 & 3 & 1 & 6 & 154 \\
\bf \#C,repro & 1 & 1 & 1 & 0 & 20 & 42 & 5 & 22 & 1 & 6 & 12 & 12 & 1 & 5 & 1 & 3 & 1 & 6 & 140 \\
\midrule
\bf diff & 0 & 0 & 0 & 1 & 0 & 8 & 0 & 3 & 0 & 2 &  0 & 0 & 0 & 0 & 0 & 0 & 0 & 0 & 14 \\
\bottomrule
\end{tabular}
}
}
\end{table}


The results of our reproducibility study on \emph{BiblioteQ} are visualized in Figure~\ref{fig_combo}.
While we are confident that we have identified the raw input data, due to liberties in the data preparation instructions, our results nevertheless deviate.

In Table~\ref{tab_biblioteq_refm}, we list the changes per revision, comparing the results of the original study against our own. 
Revision~35, where SQLite was introduced, clearly stands out.
%
%
%
%
In processing the extracted schemas (in particular,  revisions 4, 5 and 11),
we encountered small syntax errors in SQL statements, that we manually fixed to make the analysis work. 
Since we can reproduce the exact results of the original study, we may safely assume that Neamtiu et al.\ have fixed these same errors, even though they do not report this.

\paragraph{Mozilla Firefox.}

The original paper  analyzed 308 revisions of \emph{Mozilla Firefox} in a specific time interval.
From the material provided to us by the authors, we further know  
the table names in database schemas.
Unfortunately, this information was not specific enough to identify the exact revisions analyzed.
As the original version control system (CVS) has meanwhile been replaced by Mercurial, 
we inspected the \webref{https://ftp.mozilla.org/pub/vcs-archive/}{CVS archive},
the current \webref{https://github.com/mozilla/gecko-dev}{GitHub repositories},
and the Firefox \webref{https://ftp.mozilla.org/pub/firefox/releases/}{release website}.
We searched for the CVS tags mentioned by the authors, and tried to align them with these sources. Despite independent efforts by all three authors, we were not able to reliably identify the analyzed project versions.
Consequently, we are not able to report any reproducibility results.

\paragraph{Summary.}

\nop{
Some information and formulas can be found in the file:
archive/stat.xlsx (Tab: Summary)
}
\begingroup
\setlength{\tabcolsep}{2.75pt}
\begin{table}[tb]
\centering
\caption{Comparing of the total number of schema changes across projects.}
\label{tab_diff}

{\footnotesize
\resizebox{\textwidth}{!}{%
\begin{tabular}{lrrrrr}
\toprule
 & \textbf{Vienna} & \textbf{\shortstack{Monotone\\(Alt.~1: {\tt .sql})}} & \textbf{\shortstack{Monotone\\(Alt.~2: {\tt .sql}/C\texttt{++})}} & \textbf{BiblioteQ} & \textbf{\shortstack{Mozilla\\Firefox}}\\
\midrule
\textbf{Original study} & 14 & 54 & 54 & 154 & 120 \\
\textbf{Repro.\ study} & 14 & 49 & 55 & 140 & --\\
\midrule
\textbf{Abs.\ diff} & 0 & 11 & 9 & 14 & --\\
\textbf{Rel.\ diff [\%]} & 0.00 & 20.37 & 16.67 & 9.09 & --\\
\bottomrule
\end{tabular}
}
}

\end{table}
\endgroup

We summarize our results in Table~\ref{tab_diff}, which  reads as follows. For each project, we state the number of schema changes observed in the original study and in our reproducibility study. We  state the absolute difference in the results, as well as the relative difference in percent, as introduced in Section~\ref{sec:methodology}.

While we were able to exactly reproduce the results for \emph{Vienna}, we were not able to conduct the analysis  for \emph{Mozilla Firefox}. For \emph{Monotone} and \emph{BiblioteQ}, our results deviate to varying degrees.
We next discuss these effects. 

\nop{
Regarding the \textit{Monotone} project, we want to note that the reason for such a large difference is a different number of changes of the two SMOs, {\fontfamily{qcr}\selectfont CREATE TABLE} and {\fontfamily{qcr}\selectfont DROP TABLE}, as can be seen in Figure~\ref{fig_monotone}.

Finally, a comparison of changes per year is presented in Table~\ref{tab_chg_year}.

\begin{center}
\input{tab/tab_chg_year}
\end{center}
}

\section{Discussion}
\label{discussion}
Access to the raw input data,
sample code and instructions 
make project \emph{Vienna} an
almost ideal reproduction case. For the other projects, we found the data preparation instructions unspecific. For \emph{Monotone} and \emph{Mozilla Firefox}, we struggled (and in case of \emph{Mozilla Firefox} even  failed) to  locate the raw input data. Nearly a decade after the original paper has been published in 2011, code repositories have switched hosting platforms.
Therefore, \emph{a link is not enough} to unambiguously identify the raw input data, to quote from the title of a recent reproducibility study~\cite{pawlik2019}.
Further, the exact revision ranges must be clearly specified, beyond (ambiguous) dates.

The ACM reproducibility badge ``\webref{https://www.acm.org/publications/policies/artifact-review-badging}{Artefacts Available}'' requires artefacts like the raw input data 
to be available on an
 archival repository, identified by 
 a digital object identifier.
Considering our own experience, it is vital to ensure long-term access to the raw input data.
Various efforts (e.g.~\cite{10.1145/3183558}) try   to ensure long-term availability of OSS repositories. However,
 without very specific instructions on data  preparation, the reproducibility of the results remains at risk.

To quantify how much our results differ, we calculate the difference in percentage across all SMOs.
For a more fine grained assessment of the degree of reproducibility, we would require information on the exact SMOs identified in the original study. 
This motivates us to also provide the output of applying \mysqldiff in our reproducibility study in our Zenodo repository (see Section~\ref{sec:methodology}).

\section{Threats to Validity}
\label{threats}
We now turn evaluate threats to the validity of the original study,
and comment on additional threats discovered during reproduction.

\smallskip
\noindent
\emph{Threats of the Original Study.}
Three possible threats to validity are pointed out.
Firstly, missing tables in the database schema could arise from using inadequate text matching patterns.
We agree that their correctness affects result quality, especially if the pattern is used to extract schemas from code that in some versions or revisions have changed significantly. Inadequate patterns can cause missing tables, missing columns, and other issues.

Secondly, renamings are another possible source of errors. Following  usual schema history evolution techniques,
the authors consider renaming of tables and columns as a deletion followed by an addition, as implemented by \mysqldiff. Consequently, renamings cannot be correctly recognized.

Thirdly, the choice of reference systems is considered an external threat to validity. The evolution of database
schemas for applications with different characteristics might differ.

\smallskip
\noindent\emph{Threats of the Reproduction Study.}
%
The dominant threat to validity of the reproduction concerns behavior of 
\mysqldiff:
\begin{compactitem}
\item Different versions of \mysqldiff produce different output, also caused by \webref{https://www.perlmonks.org/?node_id=507294}{bugs}. 
Erroneous statements may be mistaken for actual schema changes.
%

\item Syntax errors in table declarations cause \mysqldiff to ignore
 any subsequent declarations. This error propagates, since in comparing predecessor and successor schemas, \mysqldiff will erroneously report additional SMOs, such as DROP TABLE and 
CREATE TABLE statements.

\item
 Foreign key constraints require table declarations in topological order. CREATE TABLE statements extracted from several input files require careful handling
because runtime errors may cause following inputs to be ignored.
\end{compactitem}

\mysqldiff relies on a MySQL installation, and
the \webref{https://dev.mysql.com/doc/refman/8.0/en/identifier-case-sensitivity.html}{handling} of table and column identifiers in MySQL can be case-sensitive.
%
The subject projects use lowercase table and column names, 
so this threat does not materialize. 

Finally, incorrectly selected files containing SQL statements are a threat to validity.
For instance, one individual file might be used for a specific DBMS when multiple DBMS are supported. If the schemas in different files are not properly synchronized, this leads to deviations. Carefully recording exactly which files
were analyzed is necessary.

\section{Related Work}
\label{related}

The authors of the original study~\cite{neamtiu2009,Lin:2009:CEA:1595808.1595817} analyze  on-the-fly relational schema evolution,
as well as collateral evolution of applications and databases.
Contrariwise to the object of our study~\cite{original_work}, the former was carried out \emph{manually}, and
risks differ between manual and programmatic analysis.

From the substantial body of work on empirical schema evolution studies, %
%
Curino et al.~\cite{Curino08schemaevolution} study schema evolution on MediaWiki,
and consider schema size growth, lifetime of tables and columns, and per-month revision count. They analyze schema changes at macro and micro levels. 
Moon et al.~\cite{moon2008} and Curino et al.~\cite{curino_prism2008} test the PRISM and PRIMA systems using the data set addressed in Ref.~\cite{Curino08schemaevolution}, as well as SMOs to describe schema evolution.
Qiu et al.~\cite{Qiu:2013:EAC:2491411.2491431} empirically analyze the co-evolution of relational database schemas and code in ten popular database applications.
They also discuss disadvantages of using \mysqldiff.

Pawlik et al.~\cite{pawlik2019} make a case for reproducibility in the data preparation process, and demonstrate the influence of (undocumented) decisions during data preprocessing on derived results. However, we are not aware of any reproducibility studies on schema evolution.

\section{Conclusion and Future Work}
\label{conclusion}
In this paper, we perform a reproducibility study on an analysis of the evolution of embedded database schemas. 
For one out of four real-world database applications, we obtain the exact same results; for two, we come within approx.\ 20\% of the reported changes, and fail to identify the raw input data in one case.

Our study, conducted nearly a decade after the original study, illustrates just how brittle online resources are. 
Specifically, we realize the importance of archiving the input data analyzed, since  repositories can move. This not only changes the URL, but creates further  undesirable and previously unforeseeable effects, for instance that  timestamps and tags no longer serve as identifiers.

We hope that sharing our insights, we can contribute to a more robust, collective science methodology in the data management research community.

\begin{small}
\paragraph*{Acknowledgements.}
We thank the authors of~\cite{original_work} for sharing parts of their analysis code, and their feedback on an earlier version of this report.
Stefanie Scherzinger's contribution, within the scope of project \textit{``NoSQL Schema Evolution und Big Data Migration at Scale''}, 
is funded by the Deutsche Forschungsgemeinschaft (DFG, German Research Foundation) --- grant number~385808805.
\end{small}



%
%

%
%
%
\bibliographystyle{splncs04}
\bibliography{bib}

\end{document}